\documentclass[10pt,twocolumn]{article}

\usepackage[letterpaper,margin=0.78in]{geometry}
\usepackage[T1]{fontenc}
\usepackage{lmodern}
\usepackage{microtype}
\usepackage{amsmath,amssymb}
\usepackage{graphicx}
\usepackage{booktabs}
\usepackage{caption}
\usepackage{authblk}
\usepackage{xcolor}
\usepackage[round]{natbib}
\usepackage[hidelinks]{hyperref}
\usepackage{dblfloatfix}   
\title{\bfseries A Large-Scale Database and Predictive Model of\\[2pt]
Listener-Rated Ease of Speech Understanding\\ in Commercial Hearing Aids}

\author[ ]{Andrew Sabin}
\author[ ]{Steve Taddei}
\author[ ]{Abram Bailey}
\affil[ ]{\normalsize HearAdvisor\\ \texttt{\{andy, steve, abram\}@hearadvisor.com}}
\date{June 2026}

\begin{document}
\maketitle

\begin{abstract}
\noindent
HearAdvisor aims to provide hearing-aid consumers with audio-performance metrics
and recordings that reflect real listening experience. For speech-related metrics,
HearAdvisor has historically used HASPIv2, a metric designed to predict objective
intelligibility and validated primarily under simulated distortions. Its
relationship to consumer-rated ease of understanding for commercial hearing aids
is uncertain. Here we introduce a large-scale perceptual dataset and learned metric
for listener-rated perceived benefit for speech understanding. Website visitors with
self-reported hearing loss completed a blind, MUSHRA-inspired listening test in
which they rated recordings of commercial hearing aids on a five-point ``Ease of
Understanding'' scale. The dataset contains $151{,}608$ ratings, $104{,}298$ after
quality screening, spanning $10{,}394$ binaural acoustic-manikin recordings from
$83$ commercial products across $72$ realistic acoustic scenes. To predict these
ratings, we pass aided audio and a matched clean-speech reference through a frozen
Whisper encoder, subtract their internal representations, and train a small MLP head
on the resulting difference embedding. On devices held out of training, the learned
metric substantially outperforms HASPIv2 at the scene level (overall $r=0.92$
vs.\ $0.83$; loud $=0.89$ vs.\ $0.75$; quiet $=0.79$ vs.\ $0.58$). In loud scenes,
performance reaches the split-half reliability of the listener ratings; in quiet
scenes, it approaches that ceiling. The model also responds sensibly to controlled
gain and SNR manipulations. Together, the dataset and model
provide a new way to predict listener-rated ease of speech understanding for real
commercial hearing-aid recordings.
\end{abstract}

\section{Introduction}
HearAdvisor is an independent laboratory with the goal of providing meaningful
audio performance information to hearing technology consumers. A series of
metrics, audio files, and summary scores are derived from Kemar manikin
recordings for each device and published online. Overall, the
quality of the information we provide to consumers is bounded by the extent to
which the metrics we report reflect true user preferences. The most important
metric, as rated by both consumers and clinicians (see
\citealp{sabin2023hearadvisor}), is the speech-understanding benefit across quiet
and noisy environments. Here we propose a new speech metric that is based on
(a)~our recording database of many commercial hearing aids and (b)~the
preferences of the hearing-aid consumers who visit our website.

The speech metrics on our website have historically been derived from the Hearing-Aid Speech
Perception Index, version~2 (HASPIv2; \citealp{kates2021haspi}), a widely used
model that estimates sentence-level objective intelligibility for
listeners with impaired hearing. HASPIv2 models the impaired auditory periphery
and can accurately predict objective intelligibility under a wide range of
simulated distortions. It is less clear, however, (a)~how well HASPIv2 predicts
listener performance across a span of real commercial devices and (b)~how a
change in HASPIv2 relates to a change in the user's actual perceived benefit.
Understanding both of these issues is core to HearAdvisor's goal of providing
meaningful audio performance information to consumers.

Recently, speech foundation models (large pretrained models that learn
general-purpose representations of speech, such as Automatic Speech Recognition
(ASR)) have been increasingly used as feature extractors for intelligibility
prediction. The
Clarity Prediction Challenge \citep{barker2022clarity} has shown that models
receiving the intermediate representations of ASR-based speech foundation models
can outperform HASPIv2 at predicting sentence-level intelligibility for
hearing-impaired listeners (e.g., \citealp{yu2025intrusive}, the winning
submission). However, as with HASPIv2, these evaluations did not use commercial
hearing aids and focus only on objective intelligibility.

Here we report a newly collected dataset of listener-rated \emph{ease of speech
understanding} for our large database of commercial hearing-aid recordings,
together with a model that predicts those ratings directly from audio. The
dataset comprises over $151{,}000$ ratings from website visitors with
self-reported hearing loss, who judged how easy speech was to understand on
recordings of real commercial devices across $72$ spatially-realistic acoustic
scenes. To predict these ratings, we pass
each aided recording and a matched clean-speech reference through a frozen
ASR foundation model (Whisper;
\citealp{radford2023whisper}) and subtract their internal representations to form
a ``difference'' embedding, which we map to a predicted rating with a small
trained Multilayer Perceptron (MLP) head. On devices held out of training the
model (1)~tracks human ratings closely (device-level $r=0.91$ in loud scenes,
$0.85$ in quiet), (2)~substantially outperforms HASPIv2 at predicting ratings,
and (3)~responds in sensible ways to controlled changes in audibility and
signal-to-noise ratio. Unlike the prior intelligibility models described above,
it is grounded in recordings of real commercial devices and targets the perceived, subjective
benefit reported by real hearing-aid consumers.

\section{The listening-test dataset}
\label{sec:data}
The recording pipeline, fitting decisions, and scoring methodology summarized here
are set out in full in our original whitepaper \citep{sabin2023hearadvisor} and
subsequent work (e.g., \citealp{manchaiah2024soundscore}). Here we recap only the
elements relevant to this study.

\subsection{Device recordings and acoustic scenes}
Each hearing aid is placed on a KEMAR acoustic manikin and programmed in two
ways: an \emph{initial} fit, approximating the most likely device settings (e.g.,
a manufacturer first fit), and a \emph{tuned} fit, adjusted as closely as possible
to NAL-NL2 prescriptive gain targets for experienced users \citep{keidser2011nalnl2}. For both fits, we
use the standard N3 moderately sloping audiogram \citep{bisgaard2010audiograms}.
We chose this loss because it approximates the median hearing-aid consumer. The manikin
was positioned at the center of an eight-loudspeaker ring in a purpose-built
acoustic test laboratory. Twelve everyday acoustic environments from the Ambisonic
Recordings of Typical Environments (ARTE) database were decoded to eight channels
and reproduced in the lab at the published sound levels \citep{weisser2019arte}.
Target talkers were recorded separately while listening to the corresponding
environment, to attempt to elicit the Lombard effect \citep{lombard1911}, and were
convolved with the impulse responses in the ARTE database. The speech-to-noise
ratio of each scene was set from each environment's real-world sound level,
following the level-to-SNR relationship measured for hearing-impaired listeners in
daily life \citep{wu2018snr}. Scenes are presented as one talker in front
($0$~deg), two talkers to each side ($\pm 45$~deg), or three talkers spread across
the front ($\pm 45$~deg and $0$~deg), with two talker sets per configuration. This
yields six talker configurations across twelve background environments, for a
total of $72$ scenes per device. We group the twelve backgrounds into \emph{loud}
($>70$~dB SPL; 7 backgrounds) and \emph{quiet} ($<70$~dB SPL; 5 backgrounds) scene
types, which behave differently throughout the analysis. All manikin recordings
are diffuse-field equalized to make them suitable for presentation over headphones
or speakers.

\subsection{The Blind Listening Challenge}
Human judgments of these recordings are collected through a web app on
\texttt{HearAdvisor.com} called the \emph{Blind Listening Challenge}, embedded on
each product and comparison page.

\paragraph{Coarse Calibration.}
Listeners are instructed to remove their hearing aids and listen through their
best headphones or speakers. They then perform a coarse level calibration in which
they adjust a restaurant-ambiance clip until it sounds realistic. We use a
convention where $0$~dB FS $=100$~dB SPL and assume (based on pilot testing) that on
average listeners place this around $70$~dB SPL (i.e., $-30$~dB FS). Listeners are told not to adjust their
system volume after this step, because all subsequent files follow this mapping.

\paragraph{Rating.}
The listener then rates sets of hearing aids in three scenes using a method
inspired by the MUSHRA paradigm for subjective audio quality
\citep{itu2015mushra}. On each screen the listener is presented with recordings of
six unlabeled hearing aids, each recorded in the \emph{identical} acoustic scene.
Switching between sliders changes the device under audition. For each file, the
listener uses the slider to rate how easy the speech is to understand on a
five-point ($0$--$4$) scale from ``Really Hard'' to ``Really Easy.'' We therefore
call this measure \emph{ease of speech understanding}.

\paragraph{Anchors and normalization.}
As in MUSHRA, hidden anchors are used to improve data quality. Two of the six items
on each screen are hidden anchors synthesized offline. The \emph{good} anchor meets
NAL-NL2 gain targets exactly and adds a $6$~dB improvement in signal-to-noise ratio
(SNR) via noise-only attenuation; the \emph{bad} anchor is low-pass filtered at
$1$~kHz with no SNR improvement. The anchors serve two purposes. First, they serve
as a quality-control gate: we discard a session if the listener does not rate the
good anchor at least one point above the bad anchor. Second, they serve as a
normalization scale. Within each session we linearly rescale every rating so that
the session's mean bad- and good-anchor ratings match the global mean bad- and
good-anchor ratings for the same scenes. These transformations remove per-listener
scale drift while preserving genuine differences in anchor quality across scenes.
The remaining four items on each screen are real recordings. One is the device
whose page the listener entered from; the other three are randomly chosen devices.
The three tested scenes are also chosen randomly. After the third scene the
listener is unblinded to which devices were rated and shown their average ratings.

\subsection{Participants and quality control}
\label{sec:participants}
The test is taken anonymously online, and we treat each completed session as one
participant. Across the $6{,}600$ sessions collected since the feature launched
(November 2025 -- June 2026), self-reported hearing loss is most often moderate (66\%), followed
by severe (17\%), mild (15\%), and profound (2\%). Listeners are split across
loudspeakers (41\%), headphones (34\%), and hearing aids (25\%), and together
contribute roughly $151{,}000$ individual ratings.

We then apply two filters (Table~\ref{tab:data}).
\emph{(i)~Listener inclusion.} We keep only mild, moderate, or severe hearing loss
auditioned over headphones or loudspeakers. We exclude profound loss, whose ratings
were observed to be systematically atypical. We also exclude listeners monitoring
through their own hearing aids, whose judgments are markedly less reliable and who
are in any case asked to remove them ($6{,}600 \to 4{,}878$ sessions).
\emph{(ii)~Anchor-based screening.} We discard any session in which the mean
\emph{good}-anchor rating does not exceed the mean \emph{bad}-anchor rating by at
least one point, since a listener who cannot order the two obvious anchors was not
performing the task ($4{,}878 \to 3{,}468$ sessions, $104{,}298$ ratings total).

\paragraph{Data aggregation.}
To improve data quality, we aggregate data across scenes with the same background.
Specifically, we train the model on pooled ratings across all talker configurations
that share the same background. At present, each individual recording is rated by a
median of only $6$ listeners---potentially too few for a stable target. We therefore
model at the background level, pooling the six talker configurations within each
device$\times$background$\times$fit. This aggregation raises the median number of
ratings per target to $33$. We treat those six recordings as one composite
measurement---a \emph{talker-pooled scene-level target}---for that device, fit, and
background. The pooling is justified by the high agreement across talker conditions.
The six configurations rate these targets consistently (Cronbach's $\alpha=0.90$), so
combining them sharpens the quality estimate rather than discarding real signal. Each
target is the rating-count-weighted mean of all its ratings, and the six talker
recordings enter as separate model inputs, all labelled with that shared target
score. This yields $1{,}764$ talker-pooled scene-level targets, trained from
$10{,}394$ recordings, spanning $150$ device--fits drawn from $83$ distinct products
tested.

\begin{table}[t]
  \centering\small
  \caption{Participants, quality control, and dataset.}
  \label{tab:data}
  \begin{tabular}{lr}
    \toprule
    Quantity & Value \\
    \midrule
    Listening sessions collected            & $6{,}600$ \\
    \quad after screening                   & $3{,}468$ \\
    Ratings collected                       & $151{,}608$ \\
    \quad after screening                   & $104{,}298$ \\
    Audio recordings                        & $10{,}394$ \\
    Talker-pooled scene-level targets        & $1{,}764$ \\
    Distinct products tested                & $83$ \\
    Fits per device                         & $2$ \\
    \bottomrule
  \end{tabular}
\end{table}

\section{Method}
\label{sec:method}

\subsection{Learned ease of speech understanding metric}
\begin{figure*}[b]
  \centering
  \includegraphics[width=\linewidth]{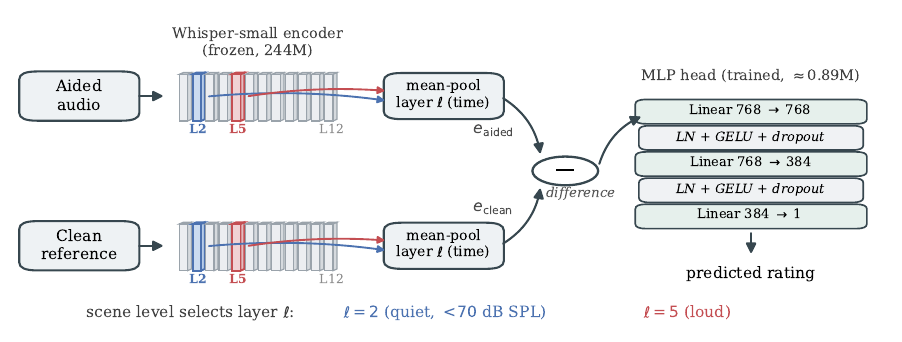}
  \caption{Model overview. Processed (aided) audio and a matched clean reference
  pass through the same frozen 12-layer Whisper-small encoder; the selected layer's
  hidden states are mean-pooled over time and differenced to form a 768-D
  difference embedding, which the trained MLP head maps to a predicted rating.
  Scene level selects the encoder layer (blue: layer 2, quiet; red: layer 5,
  loud).}
  \label{fig:diagram}
\end{figure*}

We start from a large pre-trained speech-to-text model and train only a relatively
small MLP head for our task (Fig.~\ref{fig:diagram}). We use the encoder of
Whisper-small (244M parameters; \citealp{radford2023whisper}), kept frozen. We
extract features \emph{intrusively}: the processed (aided) audio and the matched
clean reference speech (without added noise) are each downmixed to mono (the Whisper
encoder operates on a single channel) and fed separately into the encoder. We
mean-pool the encoder hidden states over time to a single
768-dimensional vector for each signal. We then compute a difference representation
by subtracting the clean-reference embedding from the aided-audio embedding,
\begin{equation}
\mathbf{x} = \mathrm{enc}_\ell(\text{aided}) - \mathrm{enc}_\ell(\text{clean}),
\end{equation}
so that the representation reflects the transformation introduced by the device
rather than the content of the utterance. The difference embedding is fed into a
small multilayer perceptron ($768\!\to\!768\!\to\!384\!\to\!1$ with layer
normalization, GELU activations and dropout; $\approx\!0.89$M trainable parameters
per head) that maps $\mathbf{x}$ to a predicted mean opinion score. The model is trained at the
talker-pooled scene level (\S\ref{sec:participants}): the six talker recordings of
a scene-level target share that target as their label, so the head sees all six as
separate examples.

We found empirically that different encoder layers $\ell$ suit different background
types: a mid-level layer (layer 5) is best for loud scenes, while an earlier layer
(layer 2) is best for quiet scenes. The metric therefore routes each file by
background level (layer 2 below $70$~dB SPL, layer 5 above), using a separately
trained head for each route.

\subsection{Training objective}
Only the two MLP heads, one for loud scenes and one for quiet, are trained; the
Whisper encoder is frozen throughout. The training loss is a weighted mean-squared
error: each target is weighted by the square root of the number of ratings behind it
(the scene-level target's total rating count), so that better-supported targets count
more. We optimize with AdamW (learning rate $10^{-3}$, weight decay
$10^{-4}$) for $200$ epochs. Because a single small head is sensitive to its random
starting point, each route's prediction is averaged over five heads trained from
different random seeds (ten heads in total, five per route).

\subsection{Baseline}
We compare against HASPIv2 \citep{kates2021haspi}, the intelligibility
metric currently used for our published speech scores. Like our model it is
reference-based (i.e., intrusive); it was computed using the standard N3 audiogram
\citep{bisgaard2010audiograms}, the same moderately sloping loss targeted by all
device fittings. To keep the comparison fair, HASPIv2 is evaluated at the same
talker-pooled scene level as our model. An important caveat is that HASPIv2
predicts \emph{objective} intelligibility, whereas our listeners rate \emph{ease of
speech understanding}, a subjective judgment related to intelligibility but
distinct from it, and one that may also reflect listening effort and overall sound
quality. The comparison quantifies how much is gained by learning from this
database rather than criticising HASPIv2 on a task it was not designed for.

\section{Experiments and results}
\label{sec:results}

\subsection{Predicting held-out devices}
\begin{figure*}[t]
  \centering
  \includegraphics[width=0.94\linewidth]{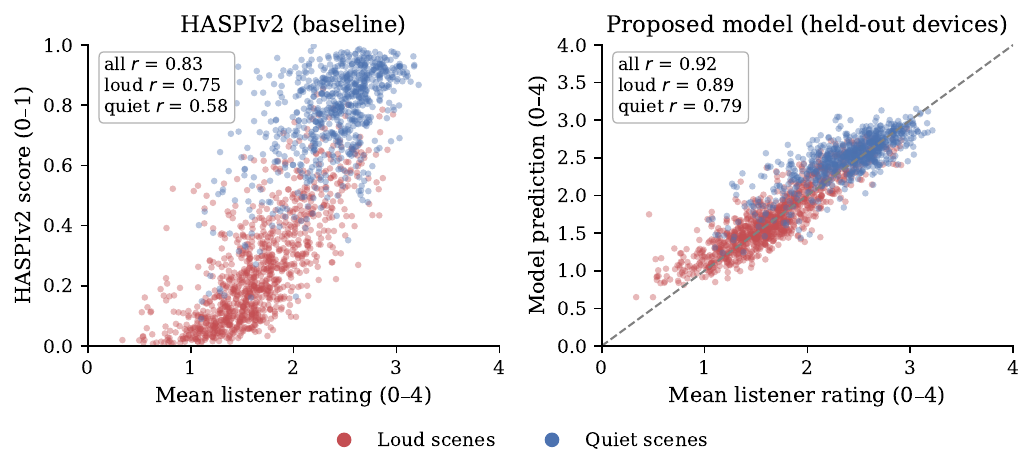}
  \caption{Each metric versus the mean listener rating of ease of speech
  understanding at the talker-pooled scene level, colored by scene type.
  \textbf{Left:} HASPIv2 correlates moderately in loud scenes and weakly in quiet
  scenes. \textbf{Right:} the learned metric, on devices held out of training, tracks
  the human ratings closely.}
  \label{fig:scatter}
\end{figure*}
We plot each metric (HASPIv2 and the proposed model) against the human mean listener
rating at the talker-pooled scene level (Fig.~\ref{fig:scatter}) defined in
\S\ref{sec:participants}. Unless noted, all correlations for our model are computed
on devices \emph{held out of training} (a device's initial and tuned variants held
out together). HASPIv2 (Fig.~\ref{fig:scatter}, left) correlates moderately with the
ratings overall ($r=0.83$) and in loud scenes ($0.75$), but more weakly in quiet
scenes ($r=0.58$). As expected, it also shows saturation at both ends of the scale
(0 and 1 proportion correct). The proposed model (Fig.~\ref{fig:scatter}, right)
tracks ratings much more closely at all levels (all $r=0.92$, loud $r=0.89$, quiet
$r=0.79$).

Because each rating average is itself a noisy estimate of true quality, the
achievable correlation is bounded by the reliability of the ratings. We estimate
this ceiling by split-half reliability: each scene-level target's ratings are split
in two, the half-means are correlated across targets, Spearman--Brown corrected,
and averaged over $1{,}000$ splits. The model effectively reaches this ceiling in
loud scenes (model $r=0.89$ vs.\ ceiling $r=0.89$), agreeing with the mean ratings
about as well as independent samples of listeners agree with each other. Quiet
scenes do not quite reach the ceiling (model $r=0.79$ vs.\ ceiling $r=0.85$).

The talker-pooled scenes are the level the model is trained on, but HearAdvisor
reports scores to consumers at two coarser levels: per \emph{fit} (a device at one
fit setting) and per \emph{device} (averaging over both fits). Aggregating the
held-out scene-level predictions and targets to these levels raises the
correlations, as averaging cancels scene-to-scene noise. At the fit level the
proposed model reaches $r=0.88$ in loud scenes and $0.81$ in quiet; at the device
level it reaches $r=0.91$ and $0.85$, respectively.

\subsection{Sensitivity to controlled manipulations}
\begin{figure}[t]
  \centering
  \includegraphics[width=\linewidth]{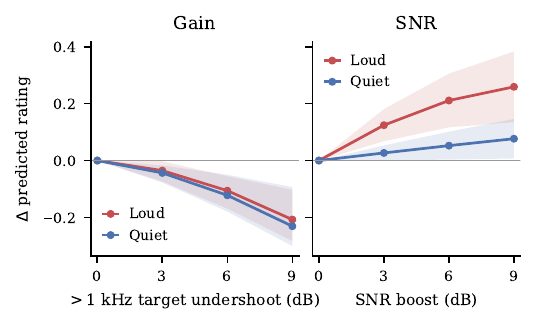}
  \caption{Change in predicted ease of speech understanding as synthetic devices
  undershoot the NAL-NL2 gain target (left) or receive an SNR boost via noise
  attenuation (right). Gain undershoot hurts both scene types; added SNR helps
  mainly in loud scenes. Bands show the spread across scenes.}
  \label{fig:sensitivity}
\end{figure}
To probe what the model has learned, we synthesized hearing-aid signals offline and
passed them through the production-trained (all-data) model while varying two
interpretable dimensions: (1)~fit-to-target gain, by intentionally undershooting the
NAL-NL2 target above $1$~kHz, and (2)~SNR, by attenuating the noise only (not the
speech). The influence of these manipulations on the model's relative output is
shown in Fig.~\ref{fig:sensitivity}. Undershooting prescribed gain targets (left) is
penalized in both scene types. Boosting SNR, however, helps much more in loud scenes
than in quiet ones (right). This is expected given that quiet scenes, unlike loud
ones, already have favorable SNRs \citep{wu2018snr} and are likely limited by
audibility rather than noise.

\section{Conclusion}
We have presented a large-scale perceptual database of listener-rated ease of
speech understanding for recordings of commercial hearing aids, and shown that
these judgments can be predicted directly from a model trained on manikin
recordings. A small MLP head on a frozen speech-recognition model predicts mean
opinion score on held-out devices well ($r=0.89$ in loud scenes, $0.79$
in quiet) and better than the objective intelligibility metric we currently use
(HASPIv2, $r=0.75$ and $0.58$, respectively). The proposed model responds sensibly
to controlled gain and SNR manipulations. Its prediction quality reaches the estimated noise
ceiling of user responses in loud scenes, and approaches it in quiet scenes.

The training data are unusually representative where it matters most to our goal.
The device sample is broad for this market: $83$ commercial products, from premium
prescription flagship products to entry-level over-the-counter devices. Every one
was recorded and rated under the same protocol. Ratings come from organic users of
our website---likely actual consumers of hearing assistance devices. Coverage of
acoustic environments is lighter. There are twelve backgrounds and six talker configurations. Expanding
this range with new recordings is an active area of our future work.

While these initial results are promising, we acknowledge other limitations.
First, we target a single standard N3 audiogram, chosen to approximate the median
hearing-aid consumer. How the proposed model generalizes to other degrees and types
of hearing loss is untested. Second, the data-collection setting is uncontrolled:
the sample is self-selected, the playback hardware varies, and the calibration is
coarse. We expect many of these factors to average out across the scale of the
dataset, but we have not directly quantified their effect.

Overall, the dataset described here captures the judgments of thousands of
hearing-aid consumers on real commercial devices in realistic acoustic scenes. The proposed
model trained on that dataset predicts those judgments about as well as the
judgments replicate across independent listener samples (\S\ref{sec:results}). We
believe this metric, once integrated into our website, brings us closer to our goal
of communicating meaningful performance information to hearing-aid consumers.


\end{document}